\documentclass{article}  
\usepackage{abstract_2e} 

\usepackage{times}
\usepackage{epsfig}
\usepackage{natbib}
\newcommand{\cotwo}{CO$_{2}$}
\newcommand{\degree}{^{\circ} }
\newcommand{\simmod}{\raise.17ex\hbox{$\scriptstyle\sim$}}

\begin{document}  


\title{Meridional Transport in the Atmospheres of Earth and Mars}


\author{A. Soto}
\affil{Southwest Research Institute, Boulder, CO, USA
(asoto@boulder.swri.edu)}


\runningtitle{}

\titlemake  

\begin{abstracttext}
\section*{Introduction}
Terrestrial atmospheres transport energy from regions of excess incoming solar radiation to regions of excess outgoing solar radiation. Generally, atmospheric energy transport occurs between equatorial and polar regions (Fig.~\ref{F:energy}). The large scale circulation, stationary eddies, and transient eddies all contribute to the meridional circulation \citep{Peixoto:1992aa}.  These are not, however, the only transport processes that can contribute to the meridional transport of energy. The Martian atmosphere has an additional mechanism for meridional transport \citep{Soto:2015aa}, and terrestrial exoplanet atmospheres may have both known and unknown meridional transport mechanisms, in addition to the large scale circulation and the eddies. An area of comparative climatology is understanding how the mechanisms for meridional transport of energy vary across the wide range of observed terrestrial planets.
 
 \begin{figure}[h]
\begin{center}
\noindent\includegraphics[width=0.45\textwidth]{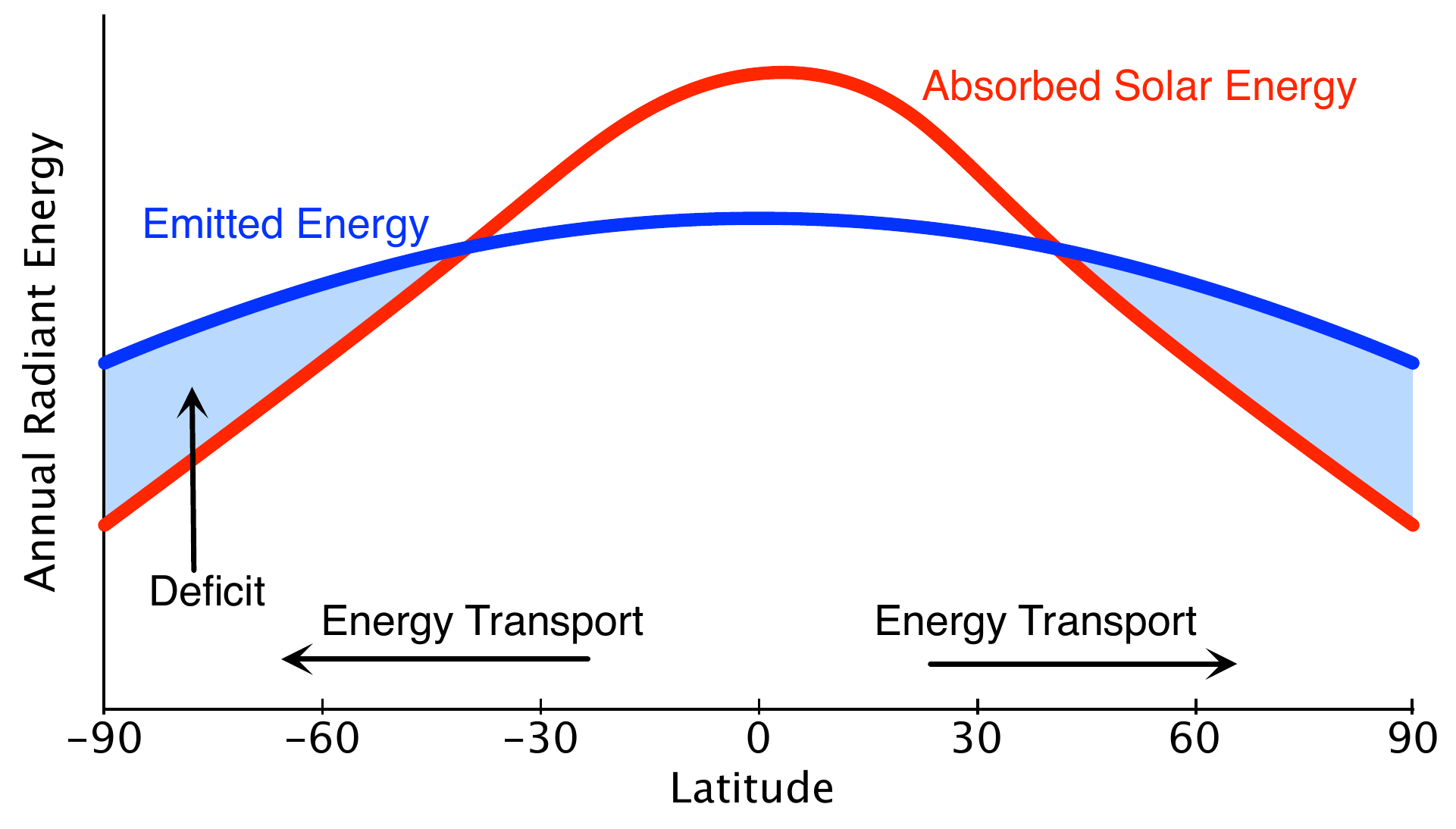}
\caption{Excess absorbed solar energy in the equatorial regions and excess emitted infrared energy in the polar regions leads to meridional energy transport, as the atmosphere strives to balance the latitudinal distribution of energy.}
\label{F:energy}
\end{center}
\end{figure}

On Earth, mean meridional circulations, stationary eddies, and transient eddies play the primary roles in the meridional transport of energy. These circulations and eddies advect the intrinsic forms of atmospheric energy, namely the internal energy, the gravitational energy, the kinetic energy, and the latent energy (due to water phase transitions). The mean meridional circulation is the main pathway for the total energy transport in Earth's tropical regions, while the eddies carry the bulk of the total energy in the extratropical regions.  This partitioning of energy transport allows the Earth's atmosphere to seasonally and annually transport net incoming radiation from the equatorial region into the polar regions \citep{Peixoto:1992aa}. 

Due to their differences in atmospheric composition and surface topography, the meridional transport of energy may exhibit very different characteristics on Earth and Mars. On Mars the drastic range in topography along with the bulk carbon dioxide's propensity to freeze in cold regions lead to a "condensation flow" that changes the partitioning of meridional energy transport \footnote{Although the phase transition of CO2 from gas to solid is properly called ÔdepositionÕ, we use the term ÔcondensationÕ in order to be consistent with previous Mars literature (e.g. \citet{Pollack:1990vn}, \citet{Haberle:1994aa}, and \citet{Forget:2013aa}).} \citep{Soto:2015aa}. When the primary atmospheric constituent, carbon dioxide, freezes out of the atmosphere, the deposition of this frozen carbon dioxide generates a flow, as the atmosphere advects carbon dioxide in order to maintain hydrostatic balance at all points. Deposition of carbon dioxide ice on the slopes of Olympus a and other large mountains generates a condensation flow that has sufficient strength to diminish the total meridional energy transported into the polar regions of Mars \citep{Soto:2015aa}.

We show numerical simulations of Earth and Mars as well as select idealized simulations (e.g., a dry, flat version of Mars and a version of Mars with a simple step-function topography). These simulations provide insight into the differences in meridional transport between Earth and Mars and the pivotal role that the exotic Martian topography plays in the atmospheric energy transport. The differences between Earth and Mars are a reminder that there may be a wide variety of meridional transport processes at work across the range of observed terrestrial planets. The traditional picture of meridional energy transport may possibly be unique to the Earth or completely common. 

\section*{Meridional Transports}
Traditionally, the zonal and vertical average of the dry static energy provides a picture of the meridional transport of energy in a terrestrial atmosphere. The dry static energy  is defined as $E = gz + c_{p}T$ where $g$ is gravity, $z$ is height, $c_{p}$ is the specific heat capacity, and $T$ is the air temperature. The dry static energy consists of the atmospheric potential energy, $gz$, and the atmospheric enthalpy, $c_{p}T$; it is the relevant energy for transport considerations \citep{Peixoto:1992aa}. 

The meridional transport of dry static energy is represented by $vE$ where $v$ is the meridional wind velocity.  We can then identify the mean circulation, $\big[\overline{v}\big]\big[\overline{E}\big]$, the stationary eddies, $\big[\overline{v}^{*}\overline{E}^{*}\big]$, and transient eddies, $\big[\overline{v'E'}\big]$, of the meridional transport, where $\overline{(\cdot)}$ is the temporal mean of the quantity $(\cdot)$ and $(\cdot)'$ is the temporal perturbation and where $[\cdot]$ is the zonal-mean of the quantity $(\cdot)$ and $(\cdot)^{*}$ is the zonal perturbation \citep{Reynolds:1895aa,Peixoto:1992aa}. The mean circulation generally drives the transport in the tropics while the eddies transport energy from the mid-latitudes to the high latitudes (Fig.~\ref{F:transport_cartoon}).

\begin{figure}[h]
\begin{center}
\noindent\includegraphics[width=0.35\textwidth]{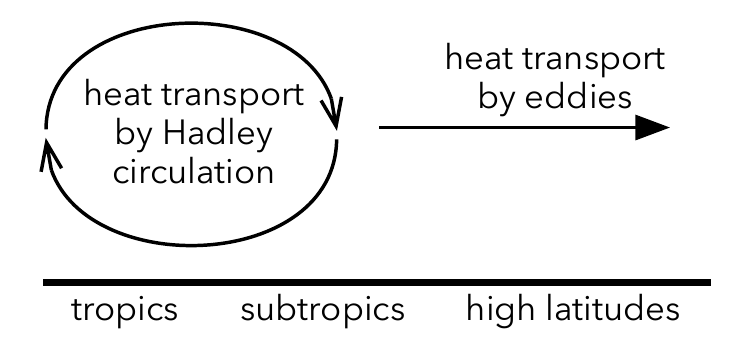}
\caption{Schematically, the Hadley cell drives transport in the tropics while eddies drive transport in the mid-latitudes.}
\label{F:transport_cartoon}
\end{center}
\end{figure}

On Earth, excess solar energy is deposited in the tropics, which generates predominantly moist convection. Although this moist convection is heterogeneously distributed and geographically confined, the zonally and annually averaged response is the generation of a Hadley circulation, where rising air in the tropics is transported poleward in the upper troposphere, before descending in the mid-latitudes. In the mid-latitudes, baroclinically-driven eddies further mix the atmospheric energy poleward. Ultimately, the excess energy in the tropics is delivered to the polar regions.

The current Martian atmosphere exhibits the same basic distribution of transport processes, except that the uplift in the Martian tropics is driven by dry convection.  If the Martian atmosphere is thicker, then under the right orbital conditions there will be an additional meridional transport mechanism at work in the Martian atmosphere. For the right combination of orbital obliquity and atmospheric mass, the Martian atmosphere undergoes large-scale, secular condensation, i.e., atmospheric collapse (e.g., \citet{Gierasch:1973fk} and \citep{McKay:1991kx}). During the atmospheric collapse process, the Martian atmosphere develops a condensation flow, which contributes, both constructively and destructively, to the meridional transport \citep{Soto:2015aa}.

\section*{Condensation Flows on Mars}
When atmospheric \cotwo{} condenses on the surface of Mars, the total mass of atmospheric \cotwo{} changes, thus changing the magnitude and distribution of the atmospheric surface pressure \citep{Tillman:1993fk,Hess:1979ys}. The atmosphere then shifts the atmospheric \cotwo{} from adjacent columns; this adjustment of the column mass due to condensation of an atmospheric constituent is known as a condensation flow \citep{Leovy:1969aa,Pollack:1981ys,Pollack:1990vn,Peixoto:1992aa}. The \cotwo{} condensation flow is the flow that remains when the meridional velocity is vertically mass-averaged  \citep{Pollack:1990vn}.  The condensation flow is calculated by taking the mass-weighted vertical average of velocity at each latitude:
\begin{equation}\label{E:cond_flow}
\big<\big[\overline{v}_{c}\big]\big> = \int^{0}_{1} \Big[\overline{(P_{s} - P_{t})v}\Big]\, \mathrm{d}\sigma
\bigg/\Big[\overline{P_{s} - P_{t}}\Big]
\end{equation}
where $v$ is the total meridional velocity at each vertical level, $P_{t}$ is a constant pressure at the top of the model, $P_{s}$ is the surface pressure, $\sigma = (P - P_{t})/(P_{s} - P_{t})$ is the vertical coordinate, $(P_{s} - P_{t})$ is the column pressure, and $\big<\cdot\big>$ indicates a vertical mean \citep{Pollack:1981ys}.  

With this condensation flow velocity, we calculate the vertically integrated meridional transport of energy due to the condensation flow, $\big<\big[\overline{v}_{c}\big]\big>\big[\overline{E}\big]$ \citep{Soto:2015aa}. With the mean circulation, $\big[\overline{v}\big]\big[\overline{E}\big]$, and the condensation flow, $ \big[\overline{v}_{c}\big]\big[\overline{E}\big]$, applied at each vertical level, we then derive the  ``overturning circulation'', $\xi$, by:
\begin{equation}
\xi = \big[\overline{v}\big]\big[\overline{E}\big] - \big[\overline{v}_{c}\big]\big[\overline{E}\big] .
\end{equation}
This method for decomposing the mean portion of the meridional transport is similar to methods previously used by \citet{Leovy:1969aa}, \citet{Pollack:1981ys}, \citet{Pollack:1990vn}, and \cite{Haberle:1993ab}.

Now we have four mechanisms involved in the meridional transport of energy:  the stationary eddies, $\big[\overline{v}^{*}\overline{E}^{*}\big]$; the transient eddies, $\big[\overline{v'E'}\big]$; the overturning circulation, $\xi$; and the condensation flow, $\big[\overline{v}_{c}\big]\big[\overline{E}\big]$. For larger atmospheric masses, the condensation flow becomes an important component of the meridional transport in the Martian atmosphere \citep{Soto:2015aa}. There is a condensation flow in the Earth's atmosphere, which is associated with the condensation of water. This condensation flow, however, is so small that it does not play a significant role in atmospheric transport in the Earth's atmosphere  \citep{Peixoto:1992aa}.

\begin{figure*}[h]
\begin{center}
\noindent\includegraphics[width=0.75\textwidth]{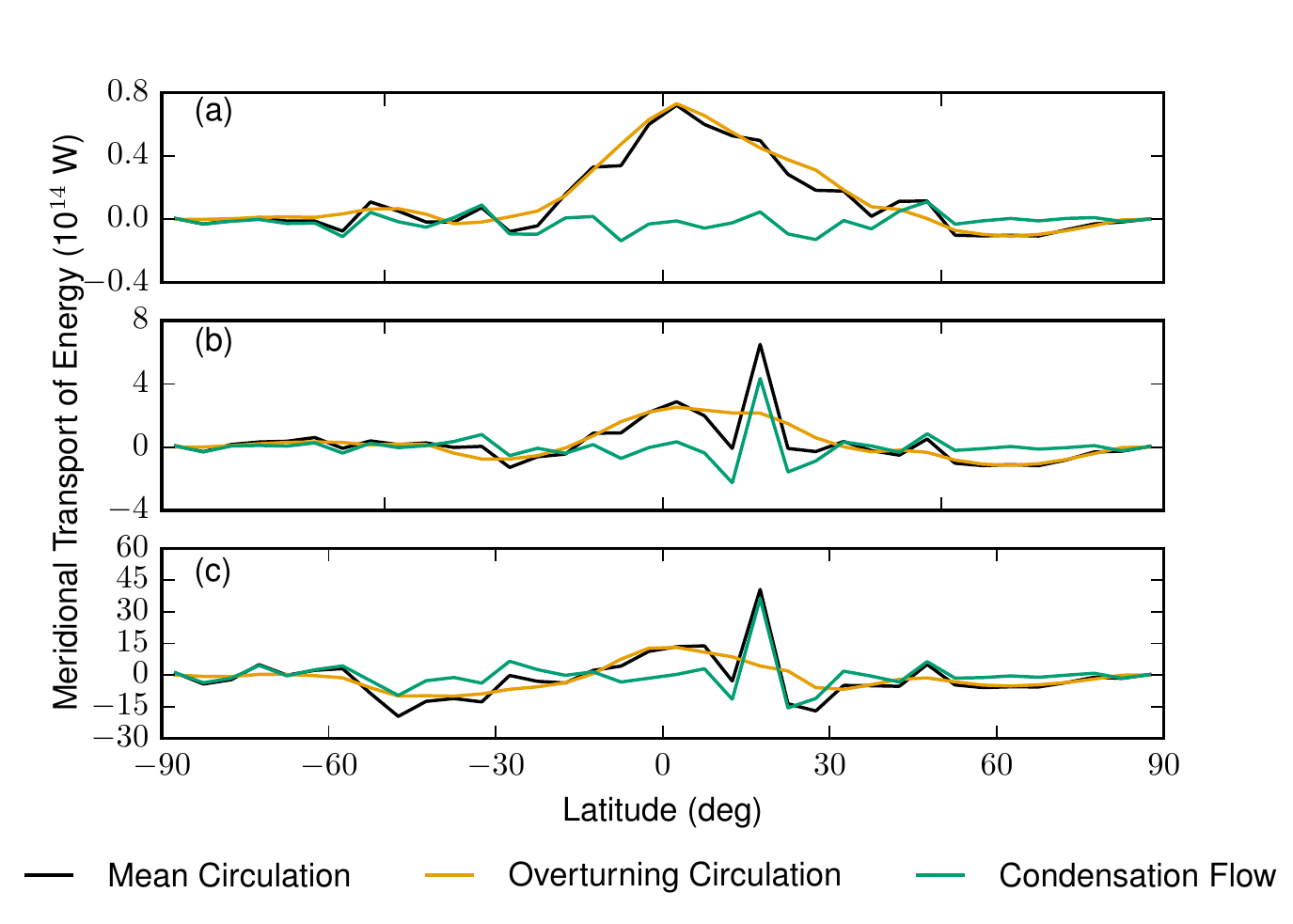}
\caption{The mean circulation, overturning circulation, and condensation flow for Mars with the current solar luminosity. This simulation used an obliquity of 25$\degree$ and an eccentricy of 0$\degree$. The meridional flows are shown for a 6 mb \cotwo{} inventory (a), a 60 mb \cotwo{} inventory (b), and a 600 mb \cotwo inventory (c). }
\label{F:cond_flow}
\end{center}
\end{figure*}

\section*{The Importance of the Condensation Flow for Mars}
On Mars, the condensation flow plays an important role in the meridional transport due to the large variation of topography.  Figure~\ref{F:cond_flow} shows the decomposition of the mean circulation into the condensation flow and the overturning circulation for three simulations of the Martian atmosphere. The simulations were created using the MarsWRF general circulation model \citep{Richardson:2007aa}. Each simulation was run for 10 years with an obliquity of 25$\degree$, an eccentricity of 0$\degree$, and the current solar luminosity. Each simulation began with a different initial \cotwo{} inventory: 6 mb in Figure~\ref{F:cond_flow}(a), 60 mb in Figure~\ref{F:cond_flow}(b), and 600 mb in Figure~\ref{F:cond_flow}(c). At the start of each simulation, all of the available \cotwo{} is in the atmosphere. After 10 years, the 6 mb simulation, Figure~\ref{F:cond_flow}(a), resembles the current Martian atmosphere, albeit with a slight difference in the \cotwo{} cycle due to the circular orbit. The 60 mb and 600 mb simulations, Figures~\ref{F:cond_flow}(b) and~\ref{F:cond_flow}(c), however, undergo atmospheric collapse.

The strong condensation flow seen in Figures~\ref{F:cond_flow}(b) and~\ref{F:cond_flow}(c) is due to the condensation of \cotwo{} onto the surface of Mars. This condensation occurs in two regions: the poles and Olympus Mons. In fact, the deposition of \cotwo ice onto Olympus Mons dominates the condensation flow and thus the mean circulation. We can confirm that the steep topography of Olympus Mons has a large influence on the latitudinal dependence of the condensation flow by looking at simulations that used a zonally averaged topography. Figure~\ref{F:zonal_flow} shows the results from three simulations that used zonally averaged topography. These three simulations are the same as those in Figure~\ref{F:cond_flow}, except for the difference in topography. Figure~\ref{F:zonal_flow} clearly shows that without  the variations in topography there is no condensation flow of any importance. This includes both the tropics and the polar regions. Thus, the existence of a condensation flow on Mars plays an important role in the meridional transport of energy, particularly once the \cotwo{} inventory is increased.

\begin{figure*}[h]
\begin{center}
\noindent\includegraphics[width=0.75\textwidth]{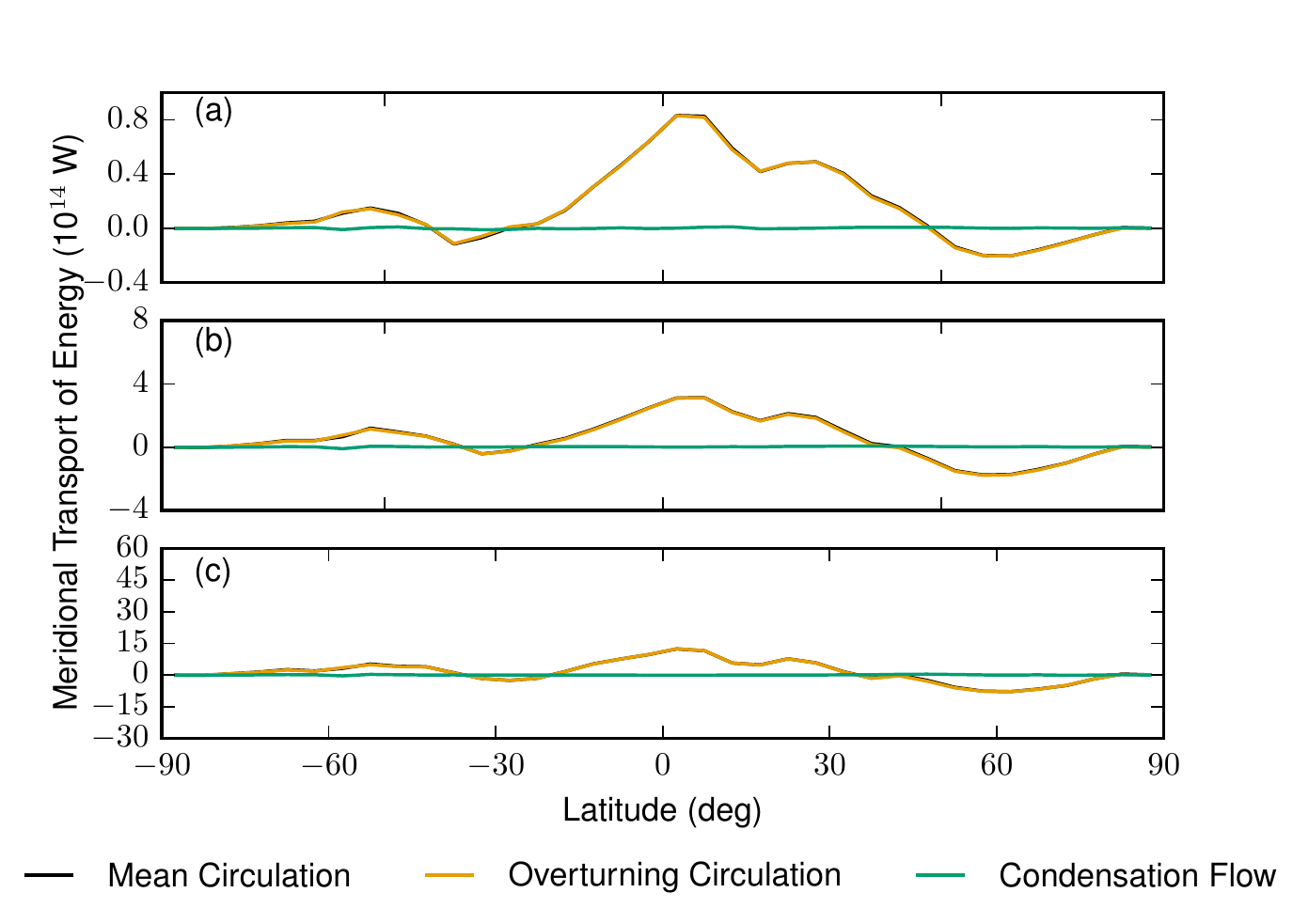}
\caption{The mean circulation, overturning circulation, and condensation flow for a Mars with a zonal averaged topography and the current solar luminosity. This simulation used an obliquity of 25$\degree$ and an eccentricity of 0$\degree$. The meridional flows are shown for a 6 mb \cotwo{} inventory (a), a 60 mb \cotwo{} inventory (b), and a 600 mb \cotwo inventory (c). }
\label{F:zonal_flow}
\end{center}
\end{figure*}

How this condensation flow affects the generation and distribution of stationary and transient eddies remains unclear. The process of atmospheric collapse drives the Martian atmosphere from an unstable state to a stable (for examples, see \citet{Gierasch:1973fk}, \citep{McKay:1991kx}, \citet{Haberle:1994aa} and \citet{Soto:2015aa}). During this process, the generation and evolution of storms may differ from the current Martian climate. Further analysis is required to complete the picture of how meridional transport and dynamics work for the range of possible Martian climates through time.

\section*{Searching for other `exotic' atmospheric processes}
This example demonstrates the importance of investigating the transport processes for each terrestrial planet that we encounter. For Mars, we have numerous atmospheric and geologic observations to guide theoretical and numerical investigations of the transport processes in the Martian atmosphere. As we look to other terrestrial planets and exoplanets, the data available to us is sparser, but we can still explore the range of possibilities through the use of numerical simulations. Atmospheric collapse likely occurs on exoplanets under a variety of conditions (e.g., \citet{Joshi:1997ly} and \citet{Wordsworth:2015aa}). For these exoplanets, understanding the condensation flow contribution to the global transport of energy may help  determine the stability conditions of exoplanet climates. 

Ultimately, this comparison between Mars and Earth highlights the possibility that non-classical atmospheric transport processes may play key roles in the overall climate dynamics of terrestrial planets. As we continue to discover terrestrial exoplanets, many with orbital and planetary characteristics drastically different from anything encountered in our solar system, we are likely to encounter other `exotic' atmospheric transport processes. We know that atmospheric collapse is a likely state for some types of terrestrial exoplanets, and thus condensation flows will likely contribute to atmospheric transport on those exoplanets. What other types of atmospheric transports may exist? Only further observations and further theoretical developments can say.

\begin{flushleft}\small{\bibliographystyle{apj}
\bibliography{/Users/asoto/Dropbox/work/science/SciencePapers/soto}}\end{flushleft}

\end{abstracttext}

\end{document}